\documentclass{article}
\usepackage{ijcai24}

\usepackage{times}
\usepackage{soul}
\usepackage{url}
\usepackage[hidelinks]{hyperref}
\usepackage[utf8]{inputenc}
\usepackage[small]{caption}
\usepackage{graphicx}
\usepackage{amsmath}
\usepackage{amsthm}
\usepackage{booktabs}
\usepackage{algorithm}
\usepackage{algorithmic}
\usepackage[switch]{lineno}
\usepackage{xcolor}
\usepackage{orcidlink}
\usepackage{svg}

\title{ProMoAI: Process Modeling with Generative AI}

\author{
Humam Kourani$^{1,2}$\orcidlink{0000-0003-2375-2152} \and
Alessandro Berti$^{1,2}$\orcidlink{0000-0002-3279-4795} \and
Daniel Schuster$^{1,2}$\orcidlink{0000-0002-6512-9580} \and
Wil M. P. van der Aalst$^{1,2}$\orcidlink{0000-0002-0955-6940}\\
\affiliations
$^1$Fraunhofer Institute for Applied Information Technology FIT, Sankt Augustin, Germany\\
$^2$RWTH Aachen University, Aachen, Germany\\
\emails
{\{humam.kourani, alessandro.berti, daniel.schuster, wil.van.der.aalst\}@fit.fraunhofer.de}
}

\begin{document}

\maketitle

\begin{abstract}
    ProMoAI is a novel tool that leverages Large Language Models (LLMs) to automatically generate process models from textual descriptions, incorporating advanced prompt engineering, error handling, and code generation techniques.
    Beyond automating the generation of complex process models, ProMoAI also supports process model optimization. Users can interact with the tool by providing feedback on the generated model, which is then used for refining the process model. ProMoAI utilizes the capabilities LLMs to offer a novel, AI-driven approach to process modeling, significantly reducing the barrier to entry for users without deep technical knowledge in process modeling. 
\end{abstract}

\section{Introduction}
Process modeling is a fundamental task in analyzing and optimizing organizational workflows. Traditional process modeling approaches require significant effort to capture workflow details and deep expertise in process modeling languages \cite{DBLP:conf/caise/IndulskaRRG09}. Such approaches often require specialized knowledge, making them inaccessible to non-experts. 

In recent years, Large Language Models (LLMs) \cite{DBLP:journals/bise/FeuerriegelHJZ24} have shown advanced capabilities in a wide array of applications, ranging from content generation to complex problem-solving. By leveraging vast amounts of data, LLMs have demonstrated remarkable advancements in understanding and generating human-like text \cite{ijcai2021p612,DBLP:conf/iclr/ZhouMHPPCB23}. 
Their ability to interpret natural language and generate coherent, contextually relevant responses makes them particularly suited for tasks that involve understanding complex textual descriptions and translating them into structured outputs.


ProMoAI introduces an innovative approach to process modeling by utilizing the capabilities of state-of-the-art LLMs, such as GPT-4 \cite{DBLP:journals/corr/abs-2303-08774}, and employing advanced prompt engineering, error handling, and code generation techniques. The primary contribution of our tool is the automation of process model generation. Furthermore, a feedback loop allows users to interactively refine the generated process models, ensuring they accurately reflect the intended processes. ProMoAI is available under \url{https://promoai.streamlit.app/}, and a demonstration video is available under {\url{https://youtu.be/t8fkx9rJmBE}}.


 \begin{figure}[!t]
    \centering        
    \includegraphics[width=0.3\textwidth]{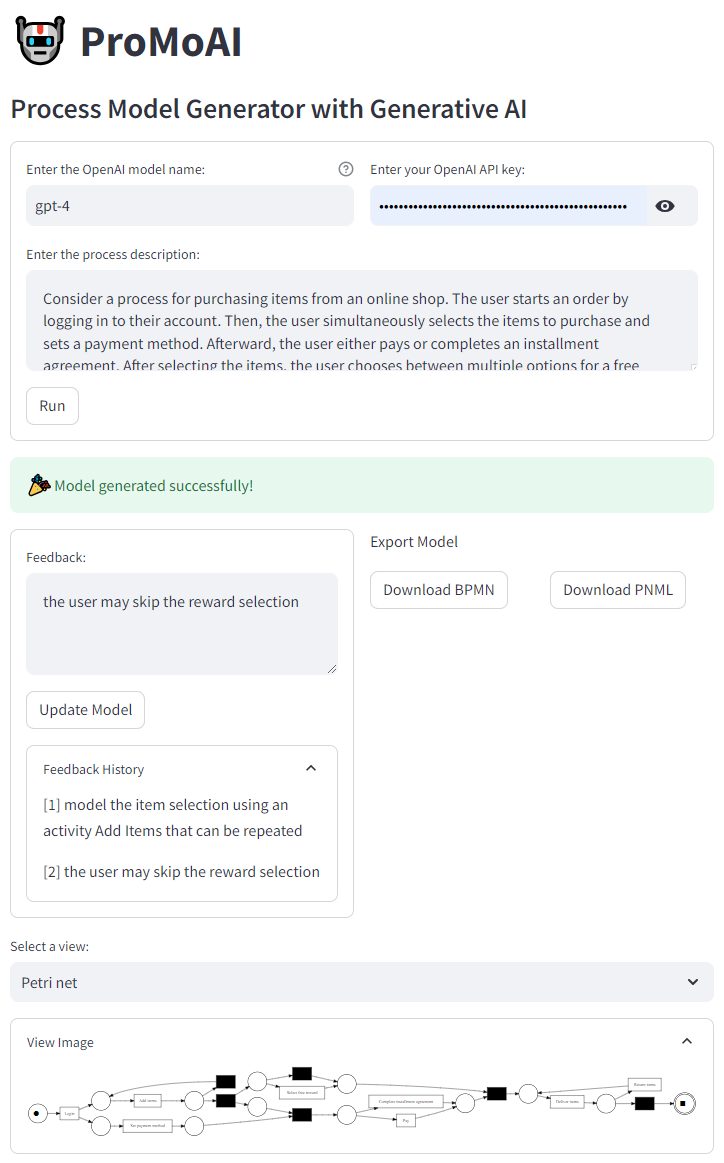}
    \caption{A screenshot of ProMoAI.}
     \label{fig:app}
 \end{figure}

\autoref{fig:app} shows a screenshot of ProMoAI. The user initially sets the LLM\footnote{Currently, only OpenAI LLMs (\url{https://openai.com/}) are supported. We are planning to support more LLMs in the future as our system's architecture is independent of the used LLM (cf. \autoref{fig:architecture}).}, enters the corresponding API key, and provides a textual description of the process. After the initial process model generation, the user can provide feedback, and the process model is updated accordingly. The generated process model can be viewed and exported in the standard modeling notation used in the fields of business process management and process mining (BPMN \cite{DBLP:books/el/15/RosingWCM15} and PNML \cite{DBLP:conf/apn/HillahKPT10}).

\section{Application Domain}
ProMoAI allows for a more intuitive and efficient creation and optimization of process models, marking a novel application of LLMs in the field of process modeling. The tool can be employed in a wide range of domains including business process management, workflow automation, and systems engineering. It is particularly beneficial in environments where rapid prototyping of process models is required to visualize and optimize workflows, making it a valuable asset for business analysts, systems engineers, and process managers.
 
\section{Related Work}
Previous works have explored utilizing natural language processing and text mining for the generation of process models (e.g., \cite{DBLP:conf/caise/FriedrichMP11} and \cite{DBLP:journals/jucs/GoncalvesSB11}). An overview of different techniques for the extraction of processes from text is provided in \cite{DBLP:conf/aiia/BellanDG20}. 


Recently, LLMs have been evaluated on business process management (BPM) tasks. In \cite{DBLP:conf/bpmds/BuschRSL23} and \cite{DBLP:conf/bpm/VidgofBM23}, the authors discuss the potentials and challenges of using LLMs for BPM tasks. In \cite{DBLP:conf/bpm/Berti0A23}, LLMs are evaluated on various process mining tasks. In \cite{DBLP:conf/bpm/KlievtsovaBKMR23}, the authors discuss the potential of LLMs in generating process models through conversational modeling. In \cite{DBLP:conf/bpm/GrohsAER23}, the authors illustrated LLMs capabilities in translating textual statements into procedural and declarative process model constraints. In \cite{DBLP:journals/emisaij/FillFK23}, the authors explored potential applications of LLMs for conceptual modeling.

\begin{figure}[!t]
    \centering        
    \includegraphics[width=0.49\textwidth]{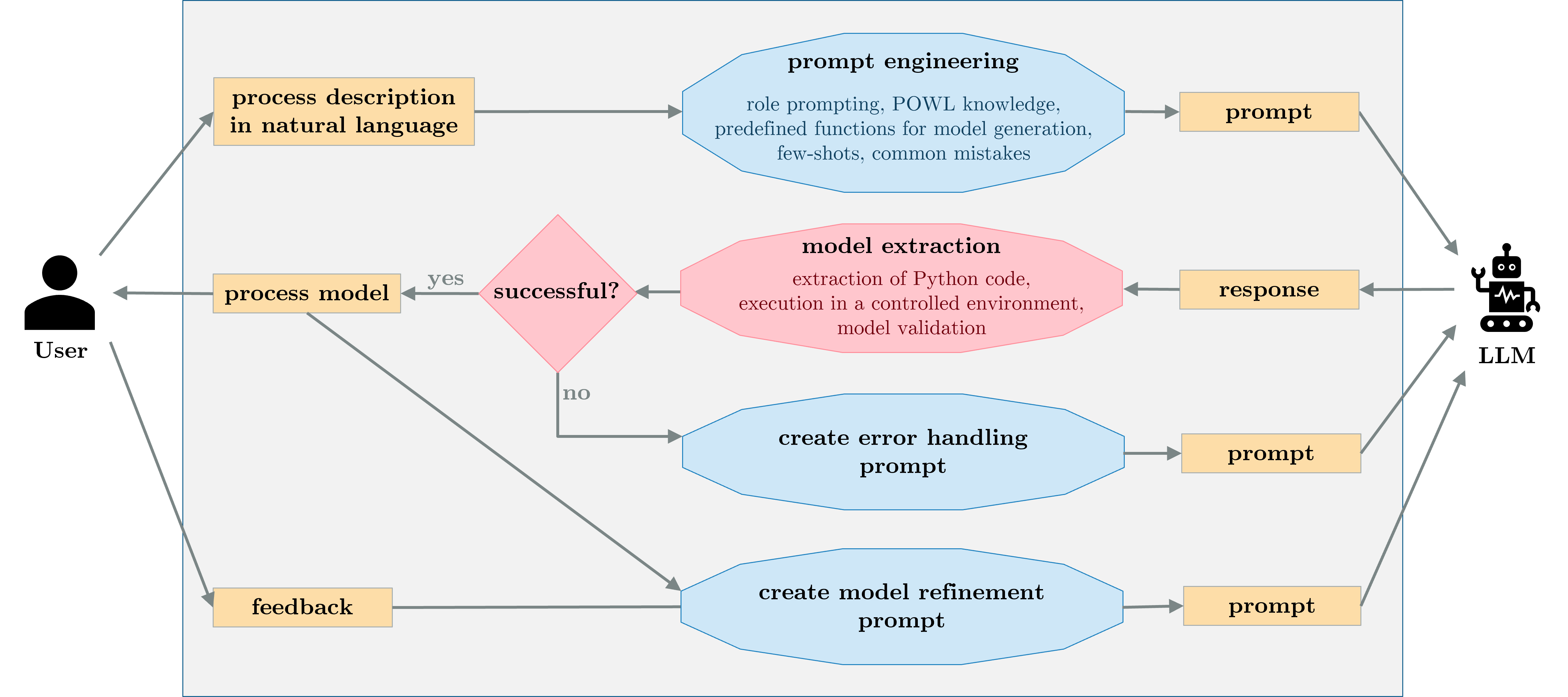}
    \caption{System architecture.}
     \label{fig:architecture}
 \end{figure}

\begin{figure*}[!t]
    \centering        
    \includegraphics[width=\textwidth]{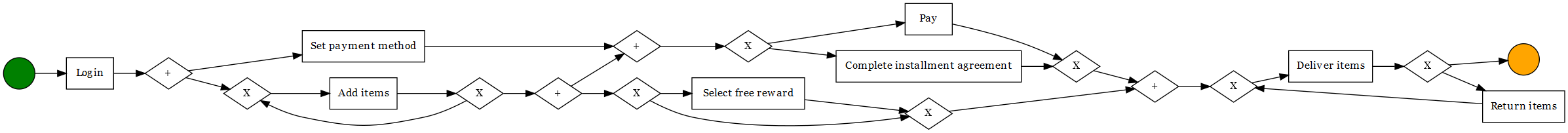}
    \caption{BPMN generated by ProMoAI for the order process using GPT-4. This model conforms to the reference model of the process, and it additionally captures the feedback requests for modeling the item selection as a loop and allowing for skipping the reward selection.}
     \label{fig:bpmn_order}
 \end{figure*}

\section{System Overview}
This section provides an overview of our methodologies.

ProMoAI implements our framework for LLM-based process modeling \cite{DBLP:conf/bpm/Humam0A24}. The high-level architecture of the underlying system is illustrated in \autoref{fig:architecture}. The user submits a textual description of a process in natural language. Our system creates a prompt that guides the LLM in generating Python code that utilizes a predefined set of functions we designed for the generation of process models (cf. \autoref{sec:prompting}). This prompt is then sent to the LLM, and based on the LLM's answer, the system attempts to generate a process model within a secure execution environment, designed to safely execute the generated code while protecting against malicious inputs or unintended operations (cf. \autoref{sec:modlegen}).
If errors are encountered during this process, an error handling prompt is sent back to the LLM, utilizing self-refinement techniques to fix the code (cf. \autoref{sec:errorhandling}). Finally, the generated model is presented to the user using standard process modeling notations (BPMN and Petri nets), with an option for exporting the process model in these formats. Additionally, the user can provide feedback, which is integrated into the system to optimize and refine the process model in an iterative refinement loop (cf. \autoref{sec:refinement}).

\subsection{Utilization of POWL}
ProMoAI employs the Partially Ordered Workflow Language (POWL) \cite{powl} for the generation of process models. POWL represents a subclass
of Petri nets that allows for the generation of block-structured models where submodels are combined to generate larger ones (i.e., a hierarchy is identified in the model). POWL and other hierarchical modeling languages are employed in the automated discovery of process models (e.g., in \cite{DBLP:conf/icpm/KouraniSA23} and \cite{DBLP:series/lnbip/Leemans22}) due to the quality guarantees they inherently provide. For example, it is not possible to generate a POWL model with dead parts that can never be reached. 

POWL allows for modeling complex, non-hierarchical process constructs while preserving the quality guarantees of hierarchical process modeling languages. POWL stands out by assuming that every task is concurrent by default unless explicitly constrained, a principle that aligns with the inherent nature of many real-world processes. This foundational assumption of concurrency facilitates the direct translation of natural language statements into POWL models, eliminating the need to specify the order of tasks unless necessary.




\subsection{Prompt Engineering Techniques} \label{sec:prompting}

ProMoAI uses sophisticated prompt engineering techniques to guide the LLM to understand the process descriptions accurately and generate the desired process models, without the extensive resource requirements and computational costs of fine-tuning \cite{DBLP:conf/ijcai/LiLTWHLB22,DBLP:conf/uai/ThangarasaGMLLD23}. 
First, we employ \emph{role prompting} \cite{DBLP:journals/corr/abs-2305-14688} by establishing a clear role for the LLM as an expert in process modeling and as a process owner who is familiar with the process context at the same time. Second, we provide a comprehensive knowledge base about POWL, offering detailed insights into its hierarchical structure, the semantics of the different POWL components, and the way POWL models are generated. Recently, LLMs have demonstrated strong capabilities in solving programming tasks and generating executable code \cite{DBLP:conf/hpec/VidanF23}. ProMoAI leverages these capabilities by instructing the LLM to transform natural language descriptions into Python code that utilizes a predefined set of functions we designed for the safe generation of POWL models.
Our strategy incorporates a \emph{least-to-most prompting} \cite{DBLP:conf/iclr/ZhouSHWS0SCBLC23} design, which outlines the task of generating a POWL model through recursive model generation, starting from simpler submodels and combining them into larger ones.
Moreover, we reinforce the LLM with instructions for self-evaluation to avoid common mistakes and ensure the process model's conformance with the process description and POWL's modeling principles.
Finally, we apply \emph{few-shot learning} \cite{DBLP:conf/nips/BrownMRSKDNSSAA20} by providing the LLM with concrete examples that illustrate how to generate POWL models starting from process descriptions (including descriptions from the PET data set \cite{DBLP:conf/bpm/BellanADGP22}).

The prompt created for our example process in \autoref{sec:example} can be found under {\url{https://github.com/humam-kourani/ProMoAI/blob/main/example_online_shop/prompt_1.txt}}.


\subsection{Process Model Generation}\label{sec:modlegen}
To ensure a high level of security and reliability, we restrict the execution of the generated Python code to a controlled environment. We implement a validation process to confirm that the generated code strictly adheres to our defined validation rules and syntax, expressly prohibiting the use of any external libraries or constructs that could introduce security risks. This methodology ensures a secure execution environment for the generation of valid process models. 

ProMoAI transforms the generated POWL models into Petri nets and BPMN models \cite{powl}. The tool provides options for viewing and exporting the generated models in these notations, which are widely recognized in the field of business process management.


\subsection{Error Handling}\label{sec:errorhandling}
ProMoAI features an error-handling mechanism that utilizes the self-evaluation and refinement capabilities of LLMs through \emph{iterative prompting} \cite{DBLP:conf/icaa2/JhaJLBVN23}. If an error occurs during the execution of the validated Python code (e.g., due to a violation of our validation rules), the system produces an informative feedback message detailing the nature of the error. This message is then incorporated into the conversation history with the LLM, enabling the LLM to refine the generated code in response to the identified error.


\subsection{Process Model Refinement}\label{sec:refinement}
A key feature of ProMoAI is its model optimization mechanism, realized through an interactive refinement loop. Users can provide feedback on the generated process models, which the system uses to generate a refinement prompt. This prompt is integrated into the conversation history with the LLM, allowing for the iterative improvement of the process model to ensure it accurately meets the user's requirements.


\section{Future Work and Extensions}
Recognizing the rapid advancement in LLM technology, ProMoAI is designed to be forward-compatible, enabling users to utilize newer OpenAI models as they become available. This flexibility ensures that ProMoAI remains at the cutting edge of AI-driven process modeling, capable of delivering improved results by leveraging the latest advancements. We also plan to support further LLMs beyond those offered by OpenAI (e.g., Google Gemini \cite{DBLP:journals/corr/abs-2312-11805}).

We are exploring the direct generation of BPMN models from textual process descriptions without the intermediate POWL representation. This will offer more flexibility in representing complex process structures and will allow us to enrich the process models with informative, context-rich annotations. However, moving away from POWL's structured guarantees requires the development of more advanced process model generation and validation techniques to maintain the quality of the generated process models.





\section{Example Application}\label{sec:example}



We applied ProMoAI using GPT-4 on the process used in \cite{powl} to describe the user's journey to order items through an online shop. This process contains typical process constructs (choice, loop, concurrency, sequence) as well as more complex non-hierarchical dependencies. After generating the initial process model, we applied two feedback comments to refine the model (cf. \autoref{fig:app}). The initial process description, the feedback comments, all generated prompts, and GPT-4's responses are available under {\url{https://github.com/humam-kourani/ProMoAI/tree/main/example_online_shop}}.


The BPMN view of the generated process model is shown in \autoref{fig:bpmn_order}. The generated model conforms to the reference process model, and it additionally integrates the feedback requests correctly into the model. This process contains complex non-hierarchical dependencies between selecting the items, setting a payment method, the reward selection, and the payment choice. As discussed in \cite{powl}, traditional hierarchical process modeling notations such as process trees \cite{DBLP:series/lnbip/Leemans22} fail to model such structures. The use of POWL enables ProMoAI to detect such structures, offering a level of flexibility and precision that traditional hierarchical process modeling languages lack.

Rerunning the application on the same input may lead to different results, e.g., GPT-4 might place the reward selection after setting the payment method. With the continuous development and advances added to LLMs, we expect future LLMs to perform better in understanding and modeling complex dependencies, leading to more consistency in the results.


\section{Conclusion}
ProMoAI represents a significant advancement in the field of process modeling, offering a user-friendly, efficient, and interactive approach to generating and refining process models. The tool allows users to describe workflows in natural language, which are then automatically translated into formal process models. By utilizing the capabilities of large language models, ProMoAI simplifies the creation of complex process models, opening up new possibilities for optimizing organizational workflows.
%


\bibliographystyle{named}
\bibliography{ijcai24}

\end{document}